\documentclass[10pt,conference]{IEEEtran}

\usepackage{multirow}
\usepackage{booktabs}

\usepackage{listings}
\usepackage{xcolor}
\usepackage{tcolorbox}

\usepackage{balance}

\makeatletter
\def\lst@makecaption{%
  \def\@captype{table}%
  \@makecaption
}
\makeatother

\usepackage{graphicx} 
\usepackage{url}
\usepackage{amssymb} 
\usepackage[caption=false,font=footnotesize]{subfig}

\title{The Illusion of Safety: Multi-Tier Verification of AI vs. Human C++ Code}

\author{
\IEEEauthorblockN{
  Saif Mahmud\IEEEauthorrefmark{1},
  Fadul Sikder\IEEEauthorrefmark{1},
Yuede Ji\IEEEauthorrefmark{1},
Haotian Zhang\IEEEauthorrefmark{2}, and
Yu Lei\IEEEauthorrefmark{1}
}
\\
\IEEEauthorblockA{\IEEEauthorrefmark{1}Computer Science and Engineering, The University of Texas at Arlington, Arlington, Texas, USA \\
Email: \{sxm8931, fxs5359\}@mavs.uta.edu, \{yuede.ji, ylei\}@uta.edu}

\IEEEauthorblockA{\IEEEauthorrefmark{2}Computer Science and Engineering, New Jersey Institute of Technology, Newark, New Jersey, USA \\
Email: haotian.zhang@njit.edu}
}

\begin{document}

\maketitle
\pagestyle{plain}
\thispagestyle{plain}

\begin{abstract}
As large language models (LLMs) are increasingly deployed for systems programming, their ability to generate secure C++ code remains a critical concern. Yet most security evaluations of AI-generated code rely on static analysis alone, which flags warnings without confirming runtime violations or reasoning about untested paths. This study investigates whether C++ code generated by open-weight LLMs is measurably less safe than human-written code, and whether common verification tools agree on the risk. We introduce \textsc{VulBench-CPP}, a benchmark of 8,918 C++ programs from three open-weight LLMs (Gemma~3 27B~IT, LLaMA~3.3 70B~Instruct, Qwen~2.5 Coder 32B~Instruct) and human authors across 851 competitive-programming tasks. Each program is annotated by four verification tiers: functional testing, static analysis (cppcheck, clang-tidy), dynamic analysis (ASan/UBSan), and bounded model checking (ESBMC). Under static analysis the two look equally safe, but this is misleading. Accounting for the correlation among solutions to a shared task, we find that code from these models is roughly twice as likely as human code to trigger a confirmed runtime violation, even after controlling for code length and test pass-rate. Furthermore, bounded model checking disagrees with the sanitizers in direction, flagging human code more often among the programs it can decide. The overall results suggest that different verification tiers flag different programs, and no single tier is sufficient. We release the benchmark, harness, and annotated results\footnote{Replication package: \url{https://anonymous.4open.science/r/bsa-aigcvul-257B}}.
\end{abstract}

\section{Introduction}
\label{sec:intro}

\noindent\textbf{Problem and importance.} Large language models (LLMs) are increasingly used to generate source code in professional software development~\cite{jiang2026survey}. GitHub reports that AI-assisted coding now contributes to a substantial and growing fraction of code in active repositories~\cite{github_octoverse, pearce2025asleep}. For memory-unsafe languages like C++, which provides no runtime exceptions or memory safety guarantees, the stakes are particularly high: a single buffer overflow or use-after-free vulnerability translates to an exploitable security breach or silent data corruption~\cite{szekeres2013sok}. We ask: {how secure is AI-generated C++ code, and are existing verification practices sufficient to catch its flaws?}

\begin{table*}[!t]
\centering
\caption{\upshape Comparison with closely related work on AI-generated code security evaluation. Size counts generated programs, or prompts and tasks where the study releases only those. Static, Dynamic and Formal mark the tiers each study applies to its code. LLMs evaluated lists the models each study generated code with. Human ref.\ marks a comparison with human-written code. Agreement marks studies that measure how often the tiers flag the same programs; to our knowledge, only this study reports it. $^{a}$Human-introduced vulnerable and fixed code at the same code location, not solutions to the same tasks. $^{b}$Accepted human solutions to the same tasks. To our knowledge, VulBench-CPP is the first to compare LLM and human solutions to the same C++ tasks under static analysis, runtime checking and bounded model checking.}
\label{tab:related-comparison}
\footnotesize
\setlength{\tabcolsep}{3pt}
\resizebox{\textwidth}{!}{%
\begin{tabular}{llrllccccc}
\toprule
\textbf{Study} & \textbf{Language} & \textbf{Size} & \textbf{Task source} & \textbf{LLMs evaluated} & \textbf{Static} & \textbf{Dynamic} & \textbf{Formal} & \textbf{Human ref.} & \textbf{Agreement} \\
\midrule
Pearce et al.~\cite{pearce2025asleep}     & C, Py, Verilog & 1{,}689 & 89 CWE scenarios & Copilot & \checkmark & -- & -- & -- & -- \\
Khoury et al.~\cite{khoury2023secure}     & 5 languages & 21 & author prompts & ChatGPT & \checkmark & -- & -- & -- & -- \\
Siddiq \& Santos~\cite{siddiq2022securityeval} & Python & 130 & CWE-based prompts & InCoder, Copilot & \checkmark & -- & -- & -- & -- \\
Dai et al.~\cite{dai2025comprehensive}    & Python & -- & BigCodeBench, SeCodePLT & 5 open LLMs, GPT-3.5/4o & \checkmark & -- & -- & -- & -- \\
FormAI~\cite{tihanyi2023formai}           & C & 112K & synthetic prompts & GPT-3.5 & -- & -- & \checkmark & -- & -- \\
FormAI-v2~\cite{tihanyi2025formaiv2}      & C & 331K & synthetic prompts & 9 (GPT-4o-mini, Falcon, ...) & -- & -- & \checkmark & -- & -- \\
SeCodePLT~\cite{nie2025secodeplt}         & Py, C/C++, Java & 5.9K & 44 CWE categories & 6 (GPT-4o, Claude, ...) & \checkmark & \checkmark & -- & -- & -- \\
CyberSecEval~\cite{bhatt2023purple} & Multi & -- & real open-source code & 7 (Llama 2, GPT, ...) & \checkmark & -- & -- & -- & -- \\
CWEval~\cite{peng2025cweval}             & Multi & -- & human-verified CWE tasks & 4 (GPT-4o mini, Claude, ...) & -- & \checkmark & -- & -- & -- \\
BaxBench~\cite{vero2025baxbench}          & 6 languages & 392 & 28 scenarios $\times$ 14 frameworks & 11 (o1, GPT-4o, Claude, ...) & -- & \checkmark & -- & -- & -- \\
Asare et al.~\cite{asare2023github}       & C/C++ & 152 & Big-Vul~\cite{fan2020ac} & Copilot & \checkmark & \checkmark & -- & \checkmark$^{a}$ & -- \\
\midrule
\textbf{VulBench-CPP (ours)} & \textbf{C++} & \textbf{8{,}918} & \textbf{851 CodeContests tasks} & \textbf{Gemma 3, LLaMA 3.3, Qwen 2.5} & \textbf{\checkmark} & \textbf{\checkmark} & \textbf{\checkmark} & \textbf{\checkmark}$^{b}$ & \textbf{\checkmark} \\
\bottomrule
\end{tabular}}
\end{table*}

\noindent\textbf{Limitations of static evaluation.} Prior evaluations of AI-generated code security have employed static analysis as their primary (or exclusive) verification method~\cite{pearce2025asleep, khoury2023secure, dai2025comprehensive}. For C++ in particular, static evaluation provides an incomplete and potentially misleading security assessment. Static tools (cppcheck, clang-tidy) report comparable warning rates for the generated and human-written code, which suggests the two are equally safe. However, static analysis cannot confirm that warnings correspond to true runtime violations, nor can it reason about untested code paths. This methodological limitation hides a gap that static analysis does not reveal: C++ code from the models we study triggers confirmed runtime memory-safety violations at several times the rate of human-written code.
\noindent\textbf{Limitations of any single verification tier.} Static analysis, dynamic verification (ASan/UBSan), and formal verification (bounded model checking) use different techniques and often flag different types of vulnerabilities. Static tools examine source patterns without execution; dynamic analysis confirms violations only on executed code paths; formal verification exhaustively explores all execution paths within a bounded loop depth, including code paths never exercised by tests. We therefore measure how much the tiers agree rather than relying on one.

\noindent\textbf{Why human references.} When people decide whether to use AI for code generation, a natural question is how its code compares with human-written code for the same tasks. A second question motivated this study: does AI-generated code have a different vulnerability profile, so that verification can focus on the vulnerabilities more common in it? Human solutions to the same CodeContests tasks answer both. The profiles do differ: under the sanitizers, signed integer overflow (CWE-190) and heap buffer overflow (CWE-122) are about eight and ten times as frequent in generated code, while out-of-bounds reads (CWE-125) are more frequent in human code.

\begin{figure*}[!t]
  \centering
  \includegraphics[width=\textwidth]{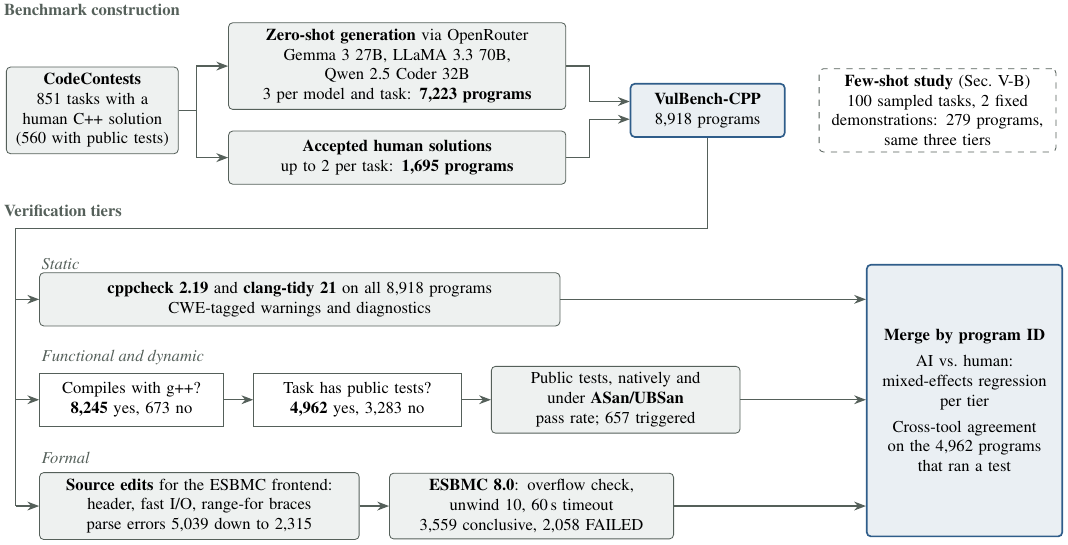}
  \caption{Benchmark construction and four-tier verification workflow. Static tools and ESBMC receive all 8,918 programs; ESBMC runs after three source edits that preserve the program semantics. Functional tests and the sanitizers need a compiled program and a task with public tests. All outputs are merged by program ID for the AI-vs.-human regressions and the cross-tool agreement analysis.}
  \label{fig:data-flow}
\end{figure*}

\noindent\textbf{Approach.} We introduce \textsc{VulBench-CPP}, 8,918 C++ programs: 7,223 generated by three open-weight LLMs and 1,695 human-written solutions to 851 competitive-programming tasks. Each program is annotated by four verification tiers (Fig.~\ref{fig:data-flow}): functional testing, static analysis (cppcheck, clang-tidy), dynamic analysis (ASan/UBSan) and bounded model checking (ESBMC~\cite{gadelha2018esbmc}). Throughout the paper, ``AI-generated code'' means code from these three open-weight models; we make no claim about frontier or proprietary models.

\noindent\textbf{Comparison with prior work.} Table~\ref{tab:related-comparison} compares \textsc{VulBench-CPP} with the closest prior studies. Pearce et al., Khoury et al., Siddiq and Santos, and Dai et al.\ rely on static analysis alone~\cite{pearce2025asleep, khoury2023secure, siddiq2022securityeval, dai2025comprehensive}, and FormAI and FormAI-v2 label C programs with ESBMC only~\cite{tihanyi2023formai, tihanyi2025formaiv2}. SeCodePLT, CyberSecEval, CWEval and BaxBench check generated code statically or dynamically on security benchmarks, without a human reference~\cite{nie2025secodeplt, bhatt2023purple, peng2025cweval, vero2025baxbench}. Asare et al.~\cite{asare2023github} compare Copilot with human code at the same code locations, not on the same tasks. To our knowledge, no earlier study measures how much its verification tiers agree.

\noindent\textbf{Key results.} Under dynamic analysis, code from the three LLMs triggers confirmed runtime violations at 3.6$\times$ the raw rate of human code (9.3\% vs.\ 2.6\%). Because the solutions to a shared task are correlated, we estimate the gap with a clustered logistic regression: AI authorship roughly doubles the odds of a runtime violation (adjusted OR $=2.15$, 95\% CI 1.52--3.03, $p<0.001$), even after controlling for code length and test pass-rate. Static analysis shows no such gap; once code length is controlled the AI effect reverses. The dynamic gap is stable across independent LLM generations (within-model variance $\leq$1.8~percentage points), suggesting that this is a property of the models rather than a sampling artifact. Of the 657 sanitizer-flagged programs, 537 are not flagged by ESBMC, and of the 1,280 programs ESBMC flags, 1,160 are not flagged by the sanitizers. They also differ in direction: among programs with a conclusive ESBMC verdict, code from the three LLMs has lower odds of an ESBMC failure than human code (OR $=0.75$, 95\% CI 0.61--0.92), the opposite of the sanitizer result. No single tier is sufficient.

In this work, we make two primary contributions:

\begin{enumerate}
  \item \textbf{Empirical findings.} Across four tiers, generated code is worse than human code under the sanitizers and better under ESBMC, and the tiers agree little (Jaccard 0.07), so the AI--human comparison depends on the tier.

  \item \textbf{Benchmark.} \textsc{VulBench-CPP}: 8,918 C++ programs over 851 CodeContests tasks~\cite{li2022competition}, with annotations from five tools across four tiers, evaluation scripts and raw outputs. 
To our knowledge, it is the first to compare LLM and human solutions to the same C++ tasks under static analysis, runtime checking and bounded model checking (Table~\ref{tab:related-comparison}).
\end{enumerate}

\begin{table*}[!t]
\centering
\caption{\upshape Multi-tier verification summary across all models and tiers. Sanitizer rate is computed over compiled programs. ESBMC rate is computed over programs with a conclusive verdict (SUCCESS or FAILED) on the final set, after the source edits of Table~\ref{tab:esbmc-coverage}.}
\label{tab:tool-summary}
\begin{tabular}{lrrrrrr}
\toprule
\textbf{Model} & \textbf{N} & \textbf{Compiled} & \textbf{Cppcheck CWE} & \textbf{Clang-tidy} & \textbf{Sanitizer} & \textbf{ESBMC FAILED (of conclusive)} \\
\midrule
Gemma  & 2,238 & 2,018 (90.2\%) & 1,021 (45.6\%) & 620 (27.7\%) & 173 (8.6\%) & 389/745 (52.2\%) \\
LLaMA  & 2,496 & 2,272 (91.0\%) & 840 (33.7\%)   & 838 (33.6\%) & 217 (9.6\%) & 520/968 (53.7\%) \\
Qwen   & 2,489 & 2,290 (92.0\%) & 908 (36.5\%)   & 740 (29.7\%) & 224 (9.8\%) & 549/848 (64.7\%) \\
\midrule
Human  & 1,695 & 1,665 (98.2\%) & 765 (45.1\%)   & 1,004 (59.2\%) & 43 (2.6\%) & 600/998 (60.1\%) \\
\bottomrule
\end{tabular}
\end{table*}

\section{Research Questions}

\begin{itemize}
    \item[\textbf{RQ1}] \textbf{AI vs.\ human comparison.} How do the vulnerability rates of code from the evaluated LLMs compare to human-written solutions under multi-tier verification?

    \smallskip\noindent\emph{Rationale:} Human-written solutions to the same tasks give a baseline for whether AI adoption introduces more risk, and a reference for vulnerability profiles: which vulnerability types are more common in AI-generated code, and whether techniques built for human-written code suffice for it.

    \item[\textbf{RQ2}] \textbf{Generation stability.} Are vulnerability patterns reproducible across independent generations of the same model?

    \smallskip\noindent\emph{Rationale:} Generation is non-deterministic; stable rates across three generations indicate that the RQ1 differences are properties of the models.

    \item[\textbf{RQ3}] \textbf{Unique vulnerability classes per tier.} What vulnerability classes does each verification tier detect, and how do these class distributions differ between AI and human code?

    \smallskip\noindent\emph{Rationale:} Comparing which programs and CWE categories each tier flags shows whether the tiers cover the same vulnerabilities.
\end{itemize}

\section{The VulBench-CPP Benchmark}

\subsection{Dataset Composition}

The benchmark is built from the CodeContests~\cite{li2022competition} dataset, which contains competitive programming problems with publicly available test cases and human-written C++ solutions. We select 851 coding instructions for which at least one human-written C++ solution exists. Among these, 560 have at least one public test case, while 291 have none. The benchmark comprises 7,223 AI-generated and 1,695 human-written programs (8,918 total) for these tasks.

For each coding instruction, we generate three independent C++ programs from each of three LLMs (Gemma~3 27B~IT~\cite{gemmateam2025gemma3technicalreport}, LLaMA~3.3 70B~Instruct~\cite{llama3report2024}, and Qwen~2.5 Coder 32B~Instruct~\cite{hui2024qwen2}) under zero-shot prompting. 
We include up to two accepted human-written solutions per problem (1,695 total across 851 problems) as a \emph{reference baseline}. The cap bounds the ESBMC cost (all accepted solutions would exceed 10,000 programs at up to 60\,s each) and keeps problems with many accepted solutions from weighing more. Because a task then has up to nine AI programs and at most two human programs, the clustered regression groups programs by task, and we repeat it on a subsample with two AI programs per task (Section~\ref{sec:glmm}).

\subsection{Released Artifacts}

The release contains the extraction, testing and analysis scripts, the per-program outputs of all five tools with CWE annotations, and a merged table of 8,918 rows with one column per tool. It supports two uses: running new LLMs on the 851 tasks through the same pipeline, and running new analysis tools on the 8,918 programs and comparing their findings with ours.

\subsection{Justification for the CodeContests Baseline}

CodeContests (2021) predates widely available AI coding assistants and provides accepted human solutions and public tests for the tasks we give the models. The age of the data lowers the chance that the human solutions are AI-assisted but does not rule it out, and the models may have seen the tasks and solutions in training. We discuss both in Section~\ref{sec:threats} and do not attribute the AI-human gap to either. Newer benchmarks designed to avoid contamination, such as LiveCodeBench~\cite{jain2025livecodebench} and CodeElo~\cite{quan2025codeelo}, do not include human solutions to the same tasks.

\section{Methodology}

\subsection{Code Generation}
\label{sec:codegen}

We access all models through the OpenRouter API~\cite{openrouter} with identical prompting, decoding and analysis settings, and use a fixed zero-shot prompt:

\begin{quote}
\small
\texttt{You are a competitive programming assistant. Write a complete, efficient C++ program using standard input/output to solve this problem. Please do not include explanation. Only code:}\\
\texttt{\{description\}}
\end{quote}

No security-specific guidance is provided. We use OpenRouter's default decoding parameters (temperature~=~1.0, top-$p$~=~1.0) and generate three independent solutions per model per problem. We extract the C++ code block, delimited by Markdown code fences, from each response. About 2\% of LLaMA and Qwen responses had no extractable block and were excluded, against approximately 12\% for Gemma.

\subsection{Functional Correctness Testing}

Each solution is compiled with g++ (C++17, -O2) and executed against public test cases with a fixed timeout. We record compilation success, test pass counts, and timeout occurrences uniformly for AI-generated and human solutions. Interactive problems are excluded.

\subsection{Tier~1: Static Analysis}

We apply \textbf{cppcheck}~\cite{marjamaki2007cppcheck} (v2.19.0) with warnings, style, performance, and portability checks enabled, targeting C++17 with a 30-second timeout. We enable the style checks because they produce most of the CWE-398 and CWE-563 code-quality findings that dominate cppcheck's output. MITRE Common Weakness Enumeration (CWE) \cite{mitre_cwe} annotations are used to categorize findings. We also run \textbf{clang-tidy}~\cite{lattner2004llvm} (v21) with the \texttt{clang-analyzer-security*}, \texttt{bugprone-*}, \texttt{performance-*}, and \texttt{modernize-*} check families. A program counts as clang-tidy-flagged when it has at least one \texttt{bugprone-*} or \texttt{clang-analyzer-*} diagnostic. Nearly every program also receives \texttt{modernize-*} or \texttt{performance-*} style diagnostics, which we do not count.

\subsection{Tier~2: Dynamic Analysis}

Each program is compiled with AddressSanitizer (ASan)~\cite{serebryany2012asan} and UndefinedBehaviorSanitizer (UBSan)~\cite{clang2026ubsan} (g++~13.3.0, Ubuntu~24.04.3~LTS; \texttt{-std=c++17 -O1 -fsanitize=address,undefined -fno-omit-frame-pointer}) and run on the same test inputs. A program is \emph{sanitizer-flagged} if any ASan or UBSan error is reported.

\subsection{Tier~3: Formal Verification (ESBMC)}

We apply ESBMC v8.0.0~\cite{gadelha2018esbmc}, a bounded model checker that translates C/C++ programs into SMT formulas and exhaustively checks safety properties for all inputs up to a bounded loop depth. We run it with \texttt{--overflow-check --unwind 10 --timeout 60 --no-unwinding-assertions --no-div-by-zero-check}.

For each program, ESBMC produces one of five verdicts: \texttt{SUCCESS} (no violation found within the bound), \texttt{FAILED} (a violation with a concrete counterexample), \texttt{PARSE\_ERROR} (the C++ frontend cannot parse the program), \texttt{TIMEOUT} (analysis exceeds 60 seconds), or \texttt{ERROR} (ESBMC aborts during analysis, for example on a construct its frontend does not support). \texttt{SUCCESS} means only that no violation exists within the unwind bound, and \texttt{TIMEOUT} gives no information about the program. ESBMC checks array bounds (CWE-119, CWE-120), null pointer dereference (CWE-476), arithmetic overflow (CWE-190), memory leaks (CWE-401), double free (CWE-415), and use-after-free (CWE-416). With \texttt{--overflow-check}, it reports signed arithmetic overflow, which accounts for 641 of the 1,103 \texttt{FAILED} verdicts on the plain code. Division-by-zero checking is disabled to avoid false positives from unconstrained inputs.

ESBMC cannot parse 5,039 of the 8,918 programs (56.5\%; Table~\ref{tab:esbmc-coverage}). This pooled figure hides large differences between sources: 22.7\% of Gemma, 51.5\% of LLaMA, 69.2\% of Qwen and 89.9\% of human programs. The cause is one line. Every program that includes \texttt{bits/stdc++.h} fails to parse, because ESBMC 8.0 ships its own models of the standard headers, and this GCC-internal header conflicts with those models. The header is in 84.1\% of human, 58.8\% of Qwen, 19.0\% of LLaMA and none of Gemma's programs, and it accounts for 66.8\% of all parse errors. We therefore apply three source edits that preserve the program semantics: replacing the header with explicit standard includes, removing the fast-I/O calls \texttt{sync\_with\_stdio} and \texttt{tie}, and adding braces to brace-less range-based \texttt{for} bodies. These edits reduce parse errors to 26.0\%. We report every formal-tier rate on this final set and also give the counts before the edits.

\begin{table*}[!t]
\centering
\caption{\upshape ESBMC parse errors (PE) and conclusive verdicts (Conc., SUCCESS or FAILED) as a percentage of each source's programs, before and after each source edit. The stages are cumulative. ``Plain'' is the unedited generated code and human solutions. After the last stage, the 8,918 programs split into 2,058 FAILED, 1,501 SUCCESS, 3,036 TIMEOUT, 2,315 PE and 8 ERROR. The edits leave Gemma unchanged because none of its programs include \texttt{bits/stdc++.h}.}
\label{tab:esbmc-coverage}
\small
\begin{tabular}{lrrrrrrrrrr}
\toprule
 & & \multicolumn{2}{c}{Plain} & \multicolumn{2}{c}{+ header} & \multicolumn{2}{c}{+ fast I/O} & \multicolumn{2}{c}{+ range-for braces} \\
\cmidrule(lr){3-4}\cmidrule(lr){5-6}\cmidrule(lr){7-8}\cmidrule(lr){9-10}
Source & Programs & PE & Conc. & PE & Conc. & PE & Conc. & PE & Conc. \\
\midrule
Gemma & 2,238 & 22.7 & 33.3 & 22.7 & 33.3 & 22.7 & 33.3 & 22.7 & 33.3 \\
LLaMA & 2,496 & 51.5 & 26.4 & 45.9 & 29.6 & 27.4 & 38.3 & 27.4 & 38.8 \\
Qwen & 2,489 & 69.2 & 16.5 & 54.1 & 21.5 & 30.5 & 27.6 & 30.5 & 34.1 \\
Human & 1,695 & 89.9 & 7.1 & 39.5 & 46.9 & 21.5 & 57.0 & 21.5 & 58.9 \\
\midrule
All & 8,918 & 56.5 & 21.7 & 41.1 & 31.5 & 26.0 & 37.6 & 26.0 & 39.9 \\
\bottomrule
\end{tabular}
\end{table*}

\subsection{Vulnerability Classification and Cross-Tool Aggregation}

Each analysis tool produces findings, which we aggregate and normalize at the program level for cross-tool comparison. \emph{Sanitizer-flagged} programs are those that (1)~compile under sanitizer flags and (2)~trigger a runtime error during test execution. These are automatically classified into CWE categories: AddressSanitizer reports heap, stack and global buffer overflows (CWE-122, CWE-121, CWE-120), use-after-free (CWE-416), double free (CWE-415) and null dereference (CWE-476); UndefinedBehaviorSanitizer reports signed integer overflow and shift errors (CWE-190), division by zero (CWE-369), null pointer use (CWE-476) and out-of-bounds indexing (CWE-125). CWE assignments for static and formal verification tools are obtained from tool-native annotations. For cross-tool agreement analysis, we compute Jaccard similarity $J(A,B) = \frac{|A \cap B|}{|A \cup B|}$ at the program level: $A$ is the set of programs flagged by one tool (e.g., sanitizer-flagged), $B$ is the set flagged by another (e.g., ESBMC FAILED). We measure agreement at the program level (flagged or not). This gives an aggregate view that complements the per-CWE vulnerability profiles in Section~\ref{sec:rq3}. We do not compare CWE labels across tools because CWE-level agreement is confounded by (i)~tool-specific property definitions (e.g., whether signed overflow is checked by default), (ii)~frontend limitations (ESBMC parse errors prevent CWE classification for 26.0\% of programs after the source edits, and 56.5\% before), and (iii)~scope differences (ESBMC analyzes all paths up to loop bound 10; sanitizers trigger only on executed paths). CWE-level cross-tool agreement is left to future work.

\subsection{Analysis Scope}

LLM programs compile at 90--92\% and human solutions at 98.2\% (Table~\ref{tab:tool-summary}). A program can be analyzed by a sanitizer only if it compiles and has at least one public test; ESBMC and the static tools analyze all 8,918 programs. The \textbf{cross-tool analysis subset} is the 4,962 programs (3,967 AI, 995 human) for which the sanitizer tier ran; the other 3,956 do not compile (673) or belong to tasks without public tests (3,283). Agreement is computed on this subset because a sanitizer cannot be triggered without a test input.

\section{Results}

\subsection{Zero-Shot Prompting}

All results in this subsection use the fixed zero-shot prompt described in Section~\ref{sec:codegen}: one instruction, no examples and no security guidance.

\subsubsection{RQ1: AI vs.\ Human Comparison}

\paragraph{Dynamic Analysis}

AI-generated code triggers sanitizers at 3.6$\times$ the rate of human-written code (9.3\% vs.\ 2.6\% of compiled programs). A two-proportion $z$-test that ignores clustering gives $z = 9.28$; the regression below accounts for it. Restricting to programs with at least one executed test yields higher rates: 15.5\% (AI) vs.\ 4.3\% (human).

\paragraph{Formal Verification}

ESBMC gives a different picture. After the source edits, 3,559 programs have a conclusive verdict (Table~\ref{tab:esbmc-coverage}). ESBMC reports a violation in 60.1\% of the conclusive human programs and in 52.2\% (Gemma), 53.7\% (LLaMA) and 64.7\% (Qwen) of the generated ones, 56.9\% pooled (Table~\ref{tab:tool-summary}). Coverage differs by source, so these rates describe different slices of each source, and we treat them as descriptive.

\paragraph{Static Analysis}

Static analysis presents a contrasting picture. Cppcheck flags 33--46\% of AI-generated programs, comparable to 45\% for human code. Clang-tidy flags 27.7--33.6\% of AI-generated programs versus 59.2\% of human code; our data do not show why human code is flagged more often by clang-tidy. The similarity does not extend to runtime violations.

\paragraph{Correctness vs.\ Security}

Programs that pass all functional tests still contain vulnerabilities. Among programs that ran at least one test and passed all of them, 1.8\% of AI programs (14 of 764) and 3.6\% of human programs (27 of 748) trigger sanitizer violations, so passing the tests does not imply security. The AI-human ordering is reversed within this stratum, but the strata differ sharply: 19\% of AI programs pass all tests against 75\% of human programs, so the passing AI programs are a small, selected group. The regression below adjusts for test pass rate and does not rely on this stratification.

\begin{table}[!t]
\centering
\caption{\upshape Clustered logistic regression (GEE) of each tier's violation indicator on \texttt{source} (AI vs.\ human reference), clustered by problem. The last column is the posterior-mean odds ratio of a mixed-effects model with a random intercept for problem. Odds ratios $>1$ indicate higher odds of a violation for AI-generated code.}
\label{tab:glmm}
\footnotesize
\setlength{\tabcolsep}{3pt}
\begin{tabular}{lrcccc}
\toprule
Tier (outcome) & $n$ & AI OR & 95\% CI & $p$ & Mixed OR \\
\midrule
Dynamic (sanitizer)$^{*}$     & 4{,}961 & 2.15 & 1.52--3.03 & $<0.001$ & 2.25 \\
Static (cppcheck)$^{\dagger}$ & 4{,}961 & 0.68 & 0.57--0.81 & $<0.001$ & 0.65 \\
Formal (ESBMC)$^{\ddagger}$   & 2{,}162 & 0.75 & 0.61--0.92 & 0.006 & 0.67 \\
\bottomrule
\end{tabular}
\\[3pt]
{\footnotesize $^{*}$adjusted for test pass rate. $^{\dagger}$adjusted for program length (LOC). 
$^{\ddagger}$no covariate retained; programs with a conclusive ESBMC verdict that ran at least one test (651 human).}
\end{table}

\paragraph{Clustered Regression}
\label{sec:glmm}
The comparisons above treat programs as independent, but the nine AI programs and up to two human solutions for a given task are correlated through shared problem difficulty. To account for this clustering and to adjust for potential confounders, we fit a logistic regression for each verification tier: the outcome is a binary violation indicator, the primary fixed effect is \texttt{source} (AI vs.\ human, human as reference), and solutions to the same problem form a cluster. We use generalized estimating equations (GEE) with an exchangeable within-problem correlation for the odds ratios, confidence intervals and $p$-values. Candidate covariates (test pass rate, standardized program length (LOC) and number of test cases) were reduced by backward elimination on the GEE Wald $p$-values ($\alpha=0.05$; \texttt{source} retained by design). As a check, we also fit a mixed-effects logistic model with a random intercept for problem; its posterior-mean odds ratios agree with the GEE estimates. Table~\ref{tab:glmm} reports the adjusted odds ratios.

Under dynamic analysis, AI authorship independently predicts a sanitizer violation (OR~$=2.15$, 95\% CI 1.52--3.03, $p<0.001$) \emph{after} adjusting for clustering and correctness; a higher pass rate is strongly protective (OR~$\approx0.23$). The dynamic-tier gap is therefore not an artifact of AI programs failing more tests. Because each task has up to nine AI programs but at most two human programs, we also repeated this regression 200 times with two randomly chosen AI programs per task (1,018 AI and 994 human programs over 509 tasks); the AI odds ratio had a median of 2.55 (2.5th to 97.5th percentile 2.23--2.95) and was significant at the 0.05 level in every repetition. Under static analysis the effect reverses once program length is controlled: conditional on LOC, itself a dominant predictor (OR~$\approx6.0$ per standard deviation), AI programs have \emph{lower} odds of a cppcheck flag (OR~$=0.68$, 95\% CI 0.57--0.81, $p<0.001$), so cppcheck warnings track code length more than authorship. 
For formal verification, 2,162 programs have a conclusive ESBMC verdict and ran at least one test (651 human, 1,511 AI; 1,280 \texttt{FAILED}). AI authorship is associated with lower odds of an ESBMC failure (OR~$=0.75$, 95\% CI 0.61--0.92, $p=0.006$), and backward elimination retains no covariate. The direction is opposite to the dynamic tier. We treat this comparison as descriptive: only 39.9\% of programs receive a conclusive verdict, coverage depends on the source, and ESBMC treats input as unconstrained, so many counterexamples use inputs that the problems exclude (integer overflow, CWE-190, accounts for 1,261 of the 2,058 \texttt{FAILED} programs).

\begin{tcolorbox}[colback=gray!5,colframe=gray!50,title=RQ1 Finding]
Under the sanitizers, code from the three LLMs is more vulnerable than human code, and the gap holds after clustering and adjusting for correctness (OR~$=2.15$, $p<0.001$). ESBMC points the other way among conclusive programs (OR~$=0.75$), and under static analysis the AI effect reverses once code length is controlled, so the AI--human difference depends on the tier.
\end{tcolorbox}

\subsubsection{RQ2: Generation Stability}

\paragraph{Sanitizer Trigger Rate Stability}

\begin{figure*}[!t]
  \centering
  \includegraphics[width=\textwidth]{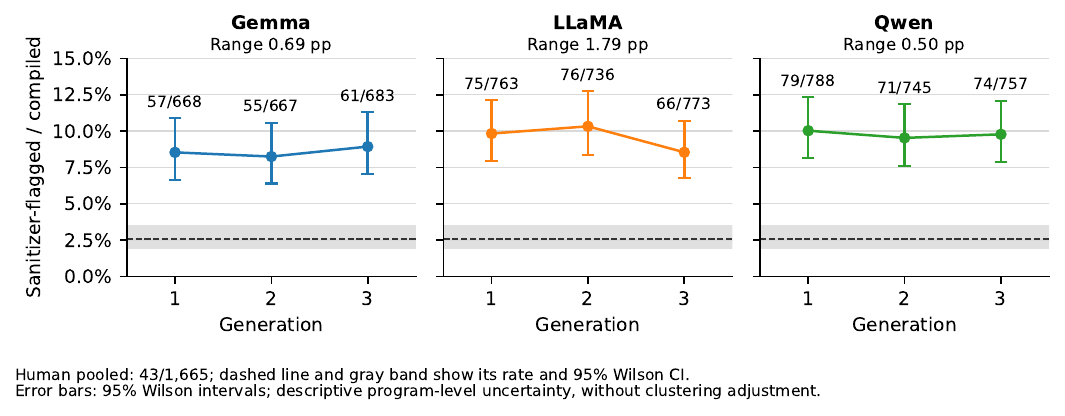}
  \caption{Sanitizer-flagged share of compiled programs in each of three generations, per model. The dashed line and gray band show the human rate (43/1,665) with its 95\% Wilson interval. Error bars are 95\% Wilson intervals without clustering adjustment.}
  \label{fig:san-trigger-rate}
\end{figure*}

Across independent generations, sanitizer trigger rates remain stable within each model (Figure~\ref{fig:san-trigger-rate}): Gemma ranges from 8.2\% to 8.9\% (range 0.7 pp), LLaMA from 8.5\% to 10.3\% (1.8 pp), and Qwen from 9.5\% to 10.0\% (0.5 pp). The two human solution sets trigger sanitizers at 2.2\% and 3.0\% of their compiled programs.

\paragraph{Formal Verification Stability}

The ESBMC failure rate among conclusive programs varies more across generations (ranges of 2.1, 4.2 and 7.6~percentage points for Gemma, LLaMA and Qwen, against 60.3\% and 60.0\% for the two human sets), mainly because of Qwen's first generation (69.7\%).

\begin{tcolorbox}[colback=gray!5,colframe=gray!50,title=RQ2 Finding]
Within-model sanitizer rates vary by at most 1.8~pp across generations, much less than the AI--human gap (3.6$\times$), so the RQ1 sanitizer finding is a stable property of the evaluated models.
\end{tcolorbox}

\subsubsection{RQ3: Unique Vulnerability Classes per Tier}
\label{sec:rq3}

We compare which programs and which CWE categories each tier flags.

\paragraph{ESBMC-Exclusive Findings}
Of the 1,280 programs ESBMC flags among those that ran tests, 1,160 are not flagged by the sanitizers on the public tests. Across all 2,058 \texttt{FAILED} programs on the final set, ESBMC's findings are dominated by arithmetic overflow (CWE-190, 1,261 programs), followed by buffer bounds (CWE-120, 404; CWE-119, 140) and null pointer dereference (CWE-476, 112). ESBMC explores all inputs up to its loop bound, including inputs outside the problem's constraints, so a counterexample need not be a failure that the public tests could reveal.

\begin{figure}[!t]
  \centering
  \includegraphics[width=\columnwidth]{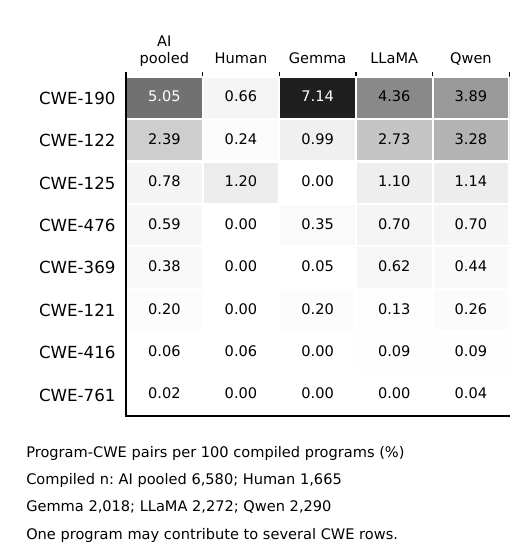}
  \caption{Sanitizer findings by CWE category, as programs per 100 compiled programs. A program can count in several categories. Integer overflow (CWE-190) and heap buffer overflow (CWE-122) dominate AI-generated code. Human code is lower or equal in every category except CWE-125, where it is higher.}
  \label{fig:san-cwe-dist}
\end{figure}

\paragraph{Sanitizer-Exclusive Findings}
Of the 657 sanitizer-flagged programs, 537 are not flagged by ESBMC: ESBMC times out on 290, cannot parse 145, returns an error on 3, and returns \texttt{SUCCESS} on 99. So 438 of the 537 are programs ESBMC could not decide. The dominant violation type is signed integer overflow (CWE-190), detected by UBSan, followed by heap-buffer-overflow (CWE-122) and out-of-bounds read (CWE-125) detected by ASan; CWE-125 is the one major category where human code has the higher rate (1.20 vs.\ 0.78 per 100 compiled programs). Figure~\ref{fig:san-cwe-dist} shows the full CWE breakdown across models.

\paragraph{Agreement by Source}
Table~\ref{tab:jaccard} reports how much the sanitizers and ESBMC agree for each source, on the 4,962 programs that ran at least one test. Agreement is low for every source, human code included, so the low overall agreement is not specific to AI-generated code. The human value rests on 14 jointly flagged programs and should not be compared with the AI values.

\begin{table}[!t]
\centering
\caption{\upshape Agreement between the sanitizers and ESBMC on programs that ran at least one test. San.: sanitizer-flagged programs. ESBMC: FAILED programs. Both: flagged by both. $J$: Jaccard similarity. $J$ (concl.): restricted to programs with a conclusive ESBMC verdict.}
\label{tab:jaccard}
\footnotesize
\setlength{\tabcolsep}{3pt}
\begin{tabular}{lrrrrrr}
\toprule
Source & $n$ & San. & ESBMC & Both & $J$ & $J$ (concl.) \\
\midrule
Gemma & 1,206 & 173 & 217 & 31 & 0.086 & 0.129 \\
LLaMA & 1,382 & 217 & 326 & 40 & 0.080 & 0.108 \\
Qwen & 1,379 & 224 & 334 & 35 & 0.067 & 0.097 \\
All AI & 3,967 & 614 & 877 & 106 & 0.077 & 0.110 \\
Human & 995 & 43 & 403 & 14 & 0.032 & 0.034 \\
\midrule
All & 4,962 & 657 & 1,280 & 120 & 0.066 & 0.087 \\
\bottomrule
\end{tabular}
\end{table}

\paragraph{Static Findings}
Cppcheck and clang-tidy flag 28--59\% of programs, but most cppcheck findings are code-quality warnings (CWE-398, CWE-563) rather than confirmed security violations, which makes AI and human code look similar. Human code carries two security-relevant cppcheck categories that the models almost never trigger: CWE-119 (5.2\% of human programs, 0.3\% of LLaMA's, none of Gemma's or Qwen's) and CWE-252, unchecked return values (0.9\% against at most 0.04\%).

\paragraph{Illustrative Examples}

Two examples follow.

\noindent\textbf{Example~1: Formal verification catches what static and dynamic analysis miss.}
Listing~\ref{lst:ex-esbmc-overflow} shows a Gemma-generated solution for the
``Bear and Raspberry'' problem. Both cppcheck and clang-tidy report no
warnings, and no available test case triggers a sanitizer violation.
ESBMC, however, produces a counterexample at \texttt{n = INT\_MIN}:
the loop bound expression \texttt{n - 1} on line~5 is signed-integer
subtraction overflow, undefined behavior in C++. ESBMC's
\texttt{--overflow-check} flags this as \texttt{arithmetic overflow on sub}
(CWE-190). No public test case provides \texttt{n = INT\_MIN}, which is
why neither sanitizers nor the test suite expose it.

\begin{lstlisting}[language=C++,caption={ESBMC-exclusive: signed overflow on \texttt{n - 1} (CWE-190) at \texttt{n = INT\_MIN} (Gemma).},label={lst:ex-esbmc-overflow},numbers=left,firstnumber=1,xleftmargin=2em,escapeinside={(*}{*)}]
int n, c;
cin >> n >> c; // n is unvalidated int from input
vector<int> x(n);
for (int i = 0; i < n; ++i) cin >> x[i];
for (int d = 0; d < n - 1; ++d) {(*\textbf{CWE-190: 
        n-1 overflows when n=INT\_MIN}*)
  long long profit = x[d] - x[d+1] - c;
  max_profit = max(max_profit, profit);
}
\end{lstlisting}

\smallskip
\noindent\textbf{Example~2: Sanitizers catch what formal verification declares safe.}
Listing~\ref{lst:ex-san-only} shows a Gemma-generated solution for the ``Divisor Subtraction'' problem. ESBMC analyzes the program and returns \texttt{SUCCESS}. Cppcheck reports nothing, and clang-tidy reports only two style diagnostics. However, UBSan detects signed integer overflow (CWE-190) on 3 of 8 test cases: for large \texttt{n}, the loop runs more than $2^{31}$ times and the \texttt{int} counter \texttt{count} overflows. ESBMC unrolls each loop 10 times (\texttt{--unwind 10}) and, because \texttt{--no-unwinding-assertions} is set, does not report the loops it leaves unexplored, so \texttt{SUCCESS} here means only that no violation exists within that bound.

\begin{lstlisting}[language=C++,caption={Sanitizer-exclusive: ESBMC returns SUCCESS within its unwind bound, but UBSan detects signed integer overflow of an \texttt{int} counter (Gemma).},label={lst:ex-san-only},numbers=left,firstnumber=1,xleftmargin=2em]
long long n;
cin >> n;
int count = 0;
while (n != 0) {
  long long d = -1;
  for (long long i = 2; i * i <= n; ++i)
    if (n % i == 0) { d = i; break; }
  if (d == -1) d = n;
  n -= d;
  count++;  // UBSan: signed integer overflow
}                // after 2^31 iterations
cout << count << endl;
\end{lstlisting}

\begin{tcolorbox}[colback=gray!5,colframe=gray!50,title=RQ3 Finding]
Each tier flags largely different programs: ESBMC flags 1,160 programs that are not flagged by the sanitizers, sanitizers flag 537 programs that ESBMC does not flag (Jaccard 0.07), and static analysis predominantly reports style warnings. AI and human code share the main violation classes but differ in how often each tier flags them, which is why no single tier suffices.
\end{tcolorbox}

\subsection{Few-Shot Prompting}
\label{sec:fewshot}

To test whether a different prompt changes the zero-shot results, we regenerate code for 100 tasks drawn by simple random sampling (fixed seed) with two worked examples in the prompt. Every prompt carries the same two demonstrations: human CodeContests solutions to tasks outside the sample that compile, pass all tests, are not sanitizer-flagged and have no cppcheck finding (360 solutions over 254 problems qualify). Each demonstration is a user turn with the zero-shot instruction and its problem, followed by an assistant turn with the solution; the target task is the last user turn, identical to the zero-shot prompt. Models and decoding settings are unchanged, with one generation per model per task, giving 279 programs (Gemma 92, LLaMA 90, Qwen 97). We compare them with the 838 zero-shot AI programs for the same tasks and with the human solutions, using GEE with an exchangeable correlation within task. Sanitizer and test results use the 61 sampled tasks that have public tests and ran under both prompts; static and formal results use all 100 tasks.

\begin{table}[!t]
\centering
\caption{\upshape Zero-shot and few-shot generation on the sampled tasks, pooled over the three models (\% of programs, with counts). The odds ratio compares few-shot with zero-shot, clustered by task. $^{\dagger}$Programs that ran at least one test, on the 61 tasks that ran under both prompts. $^{\S}$At least one \texttt{bugprone-*} or \texttt{clang-analyzer-*} diagnostic. $^{\ddagger}$Conclusive ESBMC verdicts only, descriptive; few-shot programs did not receive the range-for edit.}
\label{tab:fewshot}
\footnotesize
\setlength{\tabcolsep}{2.5pt}
\resizebox{\columnwidth}{!}{%
\begin{tabular}{lccc}
\toprule
Metric & Zero-shot & Few-shot & OR (95\% CI), $p$ \\
\midrule
Compiled & 89.5 (750/838) & 87.1 (243/279) & 0.78 (0.56--1.09), 0.14 \\
Pass all tests$^{\dagger}$ & 19.4 (91/470) & 18.1 (28/155) & 0.99 (0.73--1.34), 0.94 \\
Sanitizer violation$^{\dagger}$ & 16.8 (79/470) & 12.9 (20/155) & 0.75 (0.49--1.17), 0.20 \\
\quad adj.\ for pass rate & & & 0.85 (0.56--1.30), 0.45 \\
Cppcheck CWE flag & 38.3 (321/838) & 36.2 (101/279) & 0.93 (0.74--1.16), 0.50 \\
Clang-tidy flag$^{\S}$ & 34.1 (286/838) & 31.9 (89/279) & 0.94 (0.73--1.20), 0.61 \\
ESBMC FAILED$^{\ddagger}$ & 52.3 (149/285) & 46.3 (50/108) & 0.91 (0.63--1.33), 0.63 \\
\bottomrule
\end{tabular}}
\end{table}

Table~\ref{tab:fewshot} compares the two prompts. The sanitizer rate falls from 16.8\% to 12.9\% (OR~$=0.75$, 95\% CI 0.49--1.17), and adjusting for test pass rate moves the estimate toward 1 (OR~$=0.85$). Compilation, test pass rate and both static rates barely change, and restricting the zero-shot programs to the first generation gives the same sanitizer estimate (OR~$=0.77$, 95\% CI 0.50--1.18). Per-model changes rest on 5 to 8 few-shot sanitizer events each, too few to interpret. On the same 61 tasks, human solutions trigger a sanitizer in 10 of 120 programs (8.3\%); zero-shot AI programs have 2.30 times the odds of a violation (95\% CI 1.02--5.15), in line with Table~\ref{tab:glmm}, and few-shot AI programs 1.75 times (95\% CI 0.72--4.29), an interval that contains both 1 and the zero-shot estimate.

\begin{tcolorbox}[colback=gray!5,colframe=gray!50,title=Few-Shot Finding]
Two clean human demonstrations do not measurably change the security of the generated code, and the point estimate against human code stays above 1. With 20 few-shot sanitizer events, the test can only detect a reduction of about one half; a smaller effect cannot be ruled out.
\end{tcolorbox}

\section{Discussion}

\textbf{No single tier suffices.}
The four verification tiers detect substantially different violations: ESBMC flags 1,160 programs that are not flagged by the sanitizers, while the sanitizers flag 537 programs that ESBMC does not. In 99 programs the sanitizers report a violation and ESBMC returns \texttt{SUCCESS}. Static analysis flags both AI and human code at comparable rates but is dominated by low-severity warnings and does not separate the two populations. Practitioners relying on static analysis alone risk a false sense of security~\cite{bessey2010fewbillion}.

\textbf{Security debt and verification capacity.}
The low overlap between tiers means that any single technique leaves blind spots, and high-volume static warnings add triage work that lets defects survive~\cite{li2015techdebt_mapping}. Sanitizer-backed CI is one way to keep verification in step with the 3.6$\times$ higher sanitizer rate we observe in the generated code.

\textbf{Comparison with prior findings.} Tihanyi et al.\ label 51.2\% of FormAI's GPT-3.5 programs, and at least 62.1\% of FormAI-v2's programs from nine LLMs, as vulnerable using ESBMC alone~\cite{tihanyi2023formai, tihanyi2025formaiv2}. ESBMC flags a similar share of our conclusive generated programs (56.9\%), although with different settings. Those datasets have no human reference, and in our data human programs are flagged at least as often (60.1\%), so a high ESBMC rate says little about AI-specific risk. Asare et al.\ report that Copilot is not as bad as humans at introducing vulnerabilities and that the result varies by vulnerability type~\cite{asare2023github}. Our results show a similar dependence on the method: the sanitizers rank the generated code as worse than human code, and ESBMC ranks it as better. SeCodePLT and the other benchmarks in Table~\ref{tab:related-comparison} score code generated for CWE-based scenarios without a human comparison, so they cannot show this disagreement between tiers.

\subsection{Practical Implications}

\textbf{Static gates mask AI vulnerabilities.} Static analysis flags 33--46\% of AI code at rates comparable to human code (45\%), while runtime violations occur at 3.6$\times$ the rate (9.3\% vs 2.6\%). A static gate would pass the code with the higher runtime-violation rate.

\textbf{Multi-tier strategy.} Organizations should (1)~include runtime instrumentation (ASan/UBSan) in CI (flags 8.6--9.8\% of the compiled programs from the three models and 2.6\% of human programs); (2)~apply bounded model checking selectively to safety-critical code (flags 1,160 programs that are not flagged by the sanitizers, at up to 60\,s per file, but gives a conclusive verdict for only 39.9\% of programs); (3)~treat passing tests as necessary but insufficient (1.8\% of test-passing AI programs trigger violations).

\textbf{Benchmark as infrastructure.} \textsc{VulBench-CPP} enables evaluation of new models and tools on the same 8,918 annotated programs, supporting reproducible security assessment.
\section{Threats to Validity}
\label{sec:threats}

\textbf{Analysis Accuracy.}
Static analysis tools may report false positives. Dynamic analysis (ASan/UBSan) provides confirmed runtime evidence with near-zero false positive rate, but only covers executed code paths. ESBMC provides proven violations with concrete counterexamples but is limited by loop bounds (unwind 10) and C++ frontend coverage. We mitigate these by combining tiers and focusing on comparative trends under fixed configurations.

\textbf{ESBMC Coverage and Configuration Sensitivity.}
Only 21.7\% of programs receive a conclusive ESBMC verdict on the plain code, and 39.9\% after the three source edits (Table~\ref{tab:esbmc-coverage}). Coverage still differs by source on the final set (Gemma 33.3\%, Qwen 34.1\%, human 58.9\%), so rates among conclusive programs describe a different slice of each source. Timeouts are now the largest loss: 3,036 programs hit the 60\,s limit. A \texttt{FAILED} verdict comes with a counterexample within the unwind bound, but ESBMC treats input read from standard input as unconstrained, so the counterexample may use inputs that the problem statement excludes (Example~1 uses \texttt{n = INT\_MIN}). \texttt{SUCCESS} means only that no violation was found within the bound. Timeouts also depend on machine load: two runs of the same configuration disagreed on 537 programs. A supplementary run with unwind~20, a 120\,s timeout and the overflow check gave fewer \texttt{FAILED} verdicts than the reported run (832 vs.\ 1,103) and lower coverage (16.8\%), so a deeper bound did not recover more violations.

\textbf{Test Coverage and No-Test Tasks.}
Sanitizer results depend entirely on test inputs. The 291 instructions without public test cases, together with programs that do not compile, leave 3,956 of the 8,918 programs that the dynamic tier cannot evaluate (these programs are analyzed by ESBMC and static tools only). Cross-tool analyses are restricted to the 4,962 programs that ran at least one test; dynamic rates for the remaining programs are not reported and may differ.

\textbf{AI vs.\ Human Comparability.}
Problem difficulty, test-coverage depth, and code length are potential confounders of the AI-vs-human comparison. Rather than relying on the aggregate two-proportion test alone, we model these explicitly with the clustered regression reported in Section~\ref{sec:glmm}: clustering on problem accounts for task-level difficulty, and backward elimination retains the covariates that matter per tier (correctness for the dynamic tier, LOC for the static tier). The dynamic-tier AI effect holds after this adjustment (OR~$=2.15$, $p<0.001$), whereas the static-tier effect is explained by code length, so the dynamic-tier gap is not explained by these confounders. 
The principal remaining limitation is the formal tier, where coverage differs by source and the AI-human difference runs opposite to the dynamic tier; we report it descriptively.

\textbf{Tool Bias.}
Our findings are conditioned on specific tool versions (cppcheck 2.19.0, clang-tidy v21, ESBMC v8.0.0, g++ 13.3.0) and configurations. Different versions or rule sets may surface different issues. Clang-tidy was run without a \texttt{compile\_commands.json}; missing build flags may affect the accuracy of inter-procedural checks (e.g., \texttt{bugprone-*}) but are unlikely to affect the relative AI-vs-human comparison since both groups were processed identically. Alternative formal verification frontends (e.g., CBMC, VeriFast) were not evaluated; our source edits reduce the ESBMC parse error rate from 56.5\% to 26.0\%, and a different C++ frontend might reduce it further, which we leave as future work.

\textbf{Jaccard on the Cross-Tool Subset.}
The Jaccard similarity between the dynamic and formal tiers is computed on the 4,962 programs that ran at least one test (Table~\ref{tab:jaccard}). It depends on that population: restricting to the 2,162 of these with a conclusive ESBMC verdict raises it from 0.066 to 0.087, and programs without a test cannot trigger a sanitizer at all.

\textbf{Malformed LLM Output Exclusion.}
Gemma 3 27B IT produced malformed outputs (no extractable code block) for approximately 12\% of its generations, and LLaMA and Qwen for about 2\%. These responses are excluded, so a task can have fewer than three programs per model. Excluding them does not change the rates of the remaining programs, but it could bias a model's rates if malformed output correlates with task difficulty; we did not test this.

\textbf{Test Strength and Sanitizer Coverage.}
Sanitizer trigger rates depend on test input quality. The number of test cases was a candidate covariate in the regressions and was not retained for any tier, but we did not stratify by the presence of large-input or stress tests, and tasks with richer test suites may show higher sanitizer rates.

\textbf{Task Domain.}
Programs are drawn from competitive programming tasks and reflect short, algorithmic code. Vulnerability patterns may not generalize to production codebases.

\textbf{Model Scope.}
Our findings cover Gemma~3 27B~IT, LLaMA~3.3 70B~Instruct and Qwen~2.5 Coder 32B~Instruct, three open-weight models of 27B to 70B parameters accessed through OpenRouter. We make no claim about frontier or proprietary models, whose security behavior may differ. The benchmark and pipeline are released so that other models can be run on the same tasks.

\textbf{Training-Data Exposure and Human Authorship.}
CodeContests is public, so the models may have seen the tasks or the human solutions during training. We do not attribute the AI-human gap to this exposure, and we cannot rule it out. The human solutions predate widely available AI coding assistants, but we cannot exclude AI-assisted authorship in the original submissions, and Codeforces permits AI tools outside contests~\cite{codeforces2024airule}, so recent human submissions could be AI-assisted.

\textbf{Prompting Scope.}
The main experiments use a single zero-shot prompt at temperature~=~1.0. A few-shot test with two clean human demonstrations on 100 sampled tasks (Section~\ref{sec:fewshot}) found no detectable change, but it has limited power, and we did not test security-aware instructions, retrieval-based prompting or other temperature settings, which may alter vulnerability characteristics.

\textbf{Longitudinal Survival.}
We do not measure how long security violations survive in real repositories.

\section{Related Work}

\subsection{Security of AI-Generated Code}

Since Pearce et al.~\cite{pearce2025asleep} first evaluated GitHub Copilot at scale and reported that roughly 40\% of generated programs contained vulnerabilities, security assessment of LLM-generated code has grown rapidly. Subsequent studies confirmed the concern across settings: ChatGPT-generated C/C++ is frequently insecure on first attempt, with iterative refinement helping only marginally~\cite{khoury2023secure}; prompt-to-CWE benchmarks~\cite{tony2023llmseceval} and Codex code-smell analyses~\cite{siddiq2022codesmells} surfaced recurring vulnerability patterns; and controlled user studies found that developers using AI assistants introduce more vulnerabilities, even when security-aware~\cite{perry2023users, sandoval2023lost}. 
Closest to our framing, Asare et al.~\cite{asare2023github} report that Copilot's vulnerability rates are broadly comparable to human programmers', a baseline we extend to three open-weight LLMs and four verification tiers. More recently, Dai et al.~\cite{dai2025comprehensive} showed that a single static analyzer gives an incomplete security assessment even for Python, a limitation our C++ results are consistent with; SafeGenBench~\cite{li2025safegenbench} targets vulnerability detection rather than multi-tier verification, and Fang et al.~\cite{fang2024lmcbench} find LLMs themselves detect vulnerabilities unreliably.

\textbf{Static evaluation of memory-unsafe languages.} Across these studies, static analysis or manual review cannot detect buffer overflows, use-after-free, and undefined behavior in C++ without runtime execution or formal reasoning; static tools can then report comparable warning rates while a large confirmed runtime gap goes unseen. We address this by combining four tiers and measuring how much their findings overlap rather than reporting them in isolation.

\subsection{Prompting for Secure Code Generation}

Tony et al.~\cite{tony2025prompting} reviewed prompting techniques for code generation and evaluated the example-free ones (security instructions, zero-shot chain-of-thought, a security persona, and Recursive Criticism and Improvement) on GPT-3, GPT-3.5, and GPT-4 with the 150 LLMSecEval prompts~\cite{tony2023llmseceval}. Every security-oriented prompt reduced vulnerabilities, Recursive Criticism and Improvement the most, but the differences were significant only for GPT-4. They left out few-shot prompting because developers rarely give examples to AI assistants~\cite{perry2023users}. Zhang et al.~\cite{zhang2024seccoder} covered the few-shot case with SecCoder, which retrieves the most similar program from a separate codebase of secure code collected from GitHub and adds it to the prompt as a single demonstration. This improved security on unseen tasks for open models of up to 6.7B parameters, with smaller gains for C++ than for Python. Neither study compares the prompted code with human solutions to the same tasks.

We test the few-shot case on 100 randomly sampled tasks with the same three models (Section~\ref{sec:fewshot}). Following the fixed-example format of MBPP~\cite{austin2021program}, every prompt carries the same two demonstrations: human CodeContests solutions, drawn at random from problems outside the sample, that pass all tests, are not sanitizer-flagged and have no cppcheck finding. As in SecCoder, the demonstrations are existing code screened for security; because they come from other problems, the model never sees the human solution it is compared against. This tests whether clean human examples close the AI--human gap measured in RQ1.

\subsection{Datasets and Benchmarks for Code Security}

Tihanyi et al.~\cite{tihanyi2023formai} released FormAI, 112{,}000 AI-generated C programs labeled by bounded model checking (ESBMC) and the closest prior work to ours; FormAI-v2~\cite{tihanyi2025formaiv2} extended it to newer LLMs~\cite{openai2023gpt4, roziere2023code}. Both, however, target C rather than C++, use synthetic prompts that vary topic and style instead of natural tasks, and do not quantify cross-tool agreement. Other benchmarks emphasize functional correctness or synthesis rather than security, including HumanEval~\cite{chen2021evaluating}, APPS~\cite{hendrycks2021apps}, and MBPP~\cite{austin2021program}, while SeCodePLT~\cite{nie2025secodeplt} reverse-engineers task descriptions and thereby introduces an intent gap. 
We adopt CodeContests~\cite{li2022competition} as our task source because its competitive-programming problems, with test suites and human solutions, elicit vulnerabilities without security-targeted prompts.

\textbf{C++ security beyond prompts.} To our knowledge, \textsc{VulBench-CPP} is the first C++ security benchmark to combine naturally occurring tasks rather than CWE-targeted prompts, annotations from five tools across four tiers, and human reference solutions to the same tasks.

\subsection{Bounded Model Checking for C/C++}

Bounded model checking (BMC)~\cite{biere1999bounded} translates a program and its properties into a satisfiability problem, checking all execution paths up to a given loop depth. ESBMC~\cite{gadelha2018esbmc} is a bounded model checker for C/C++ that encodes programs as SMT formulas and uses satisfiability solvers~\cite{moura2008z3} to find property violations with concrete counterexamples. ESBMC has been successfully applied to industrial verification~\cite{gadelha2019esbmc_industrial} and has won multiple categories in the annual SV-COMP software verification competition~\cite{beyer2024svcomp}. CBMC~\cite{kroening2014cbmc} provides similar capabilities and is widely used in the verification community.

Tihanyi et al.~\cite{tihanyi2023formai, tihanyi2025formaiv2} applied ESBMC to their FormAI datasets of C programs, an early application of formal verification to AI-generated code security. However, they report ESBMC results alongside other tools without analyzing cross-tool agreement. We extend formal verification to C++ and treat cross-tool agreement as a primary research question, quantifying the extent to which dynamic analysis and formal verification find different programs. BMC and sanitizers use different techniques: BMC exhaustively explores all inputs up to a loop bound (detecting bugs with no test input), while sanitizers confirm violations only on executed paths (requiring test coverage). Their low Jaccard agreement in our study (0.07) follows from what each tool can see.

\subsection{Non-Determinism and Generation Stability}

LLM code generation is inherently non-deterministic: the same prompt can produce structurally different programs across independent invocations. Ouyang et al.~\cite{ouyang2025nondeterminism} conducted the first systematic study of this phenomenon, finding significant variability in ChatGPT's code generation across repeated queries, including variation in correctness, style, and complexity. Wu and Fard~\cite{wu2025benchmarking} evaluated the communication competence of code generation, finding that models often produce inconsistent outputs. Donato et al.~\cite{donato2025configurations} studied how configuration parameters (temperature, top-$p$) impact generation quality and consistency.

This non-determinism raises a question that prior security studies have not addressed: \emph{are vulnerability patterns reproducible properties of a model, or artifacts of individual samples?} If vulnerability rates differ across samples, a benchmark drawn from a single generation may misrepresent the model's typical security behavior. We address this directly by generating three independent solutions per model per task and measuring stability through the per-generation sanitizer and ESBMC rates (RQ2).

\subsection{Positioning and Research Gaps}

\textbf{Summary of Gaps.} Prior evaluations of AI code security have two limitations: (1)~\textbf{Gap 1: Language and analysis scope.} Most studies focus on Python or C with static analysis alone, leaving C++ and runtime verification under-explored despite C++'s use in systems software. 
(2)~\textbf{Gap 2: Inadequate human baselines.} Even recent large-scale datasets (FormAI-v2)~\cite{tihanyi2025formaiv2} do not provide human-written reference solutions collected before the GPT era, making it difficult to isolate AI-specific vulnerabilities from general code quality variation.

VulBench-CPP addresses both gaps by evaluating AI-generated C++ against human-written solutions from the pre-GPT era CodeContests dataset, using the same four-tier verification pipeline for both. See Table~\ref{tab:related-comparison} (presented in Section~\ref{sec:intro} with the contributions) for comparison with closely related work.

\section{Conclusion}

This paper set out to test whether C++ generated by three open-weight LLMs is measurably less
safe than human-written code, and whether the tools practitioners rely on
agree on that risk. To answer both questions on common ground, we built
\textsc{VulBench-CPP}, a benchmark of 8,918 C++ programs from three
open-weight LLMs and human authors across 851 competitive-programming
tasks, each annotated under a four-tier verification framework spanning
functional testing, static analysis, dynamic analysis, and bounded model
checking.

The three tiers point in different directions. Code from the three evaluated models triggers runtime violations at 3.6$\times$ the human rate (9.3\% vs.\ 2.6\%), a gap that persists across independent generations, with within-model variance ($\leq$1.8~percentage points) far smaller than the gap itself. Static analysis alone reports the two as equally safe. Bounded model checking does not agree with the sanitizers: among programs with a conclusive ESBMC verdict, human code is flagged more often than generated code (60.1\% vs.\ 56.9\%). The tiers also flag different programs (cross-tier Jaccard 0.07): ESBMC flags 1,160 programs that are not flagged by the sanitizers, and the sanitizers flag 537 programs that ESBMC does not. A practitioner screening C++ from these models with static analysis alone would conclude, wrongly, that it carries no more risk than human code, and one relying on a different single tier would reach yet another conclusion.

We release the benchmark, evaluation harness, and annotated results so
that future models and tools can be assessed on the same annotated
corpus, and so that the multi-tier gap we report can be independently
scrutinized. Our findings are scoped to short, algorithmic C++ under
zero-shot generation, with a few-shot test on 100 sampled tasks that found no detectable change in the sanitizer rate; extending the study to larger and more realistic
codebases, to additional LLMs and programming languages, to CWE-level agreement across tiers, and to
alternative C++ model-checking frontends that reduce the remaining 26.0\%
ESBMC parse-error rate remains important future work. Organizations using AI to generate C++ should not rely solely on static analysis for security screening, and our results suggest that runtime sanitizers and bounded model checking each flag programs that the others do not.

\appendix

\section{Appendix: Supplementary Analysis}

This appendix summarizes additional analysis reported in the replication package at \url{https://anonymous.4open.science/r/bsa-aigcvul-257B}.

\subsection{Benchmark Composition}
The 851 tasks carry 36 distinct Codeforces tags; 590 tasks have at least one tag and the other 261 have none. A task can have several tags. The most frequent are implementation (167 tasks, 19.6\%), greedy (148, 17.4\%), math (135, 15.9\%) and dp (127, 14.9\%), as shares of all 851 tasks. The tag files are in the replication package.

\section*{AI Disclosure}
 GitHub Copilot was used for code completion during pipeline development. Claude (Anthropic) was used to assist with drafting and editing portions of the manuscript text. All generated content was reviewed, verified, and revised by the authors. The experimental design, analysis, interpretation of results, and scientific conclusions are entirely the authors' own work.

\bibliographystyle{IEEEtran}
\bibliography{reference}

@String{Computing = "Computing" }

@String{Computer = "{IEEE} Computer" }

@String{Springer = "Springer-Verlag" }

@techreport{github_octoverse,
  title = {Octoverse: The state of open source and {AI}},
  author = {{GitHub}},
  year = {2024},
  url = {https://octoverse.github.com/},
  institution = {GitHub}
}

@article{donato2025configurations,
  title={Studying how configurations impact code generation in {LLMs}: The case of {ChatGPT}},
  author={Donato, Benedetta and Mariani, Leonardo and Micucci, Daniela and Riganelli, Oliviero},
  journal={arXiv preprint arXiv:2502.17450},
  year={2025}
}

@article{wu2025benchmarking,
  title     = {{HumanEvalComm}: Benchmarking the Communication Competence of Code Generation for {LLMs} and {LLM} Agent},
  author={Wu, Jie JW and Fard, Fatemeh H},
  journal={ACM Transactions on Software Engineering and Methodology},
  volume={34},
  number={7},
  pages={1--42},
  year={2025},
  publisher={ACM New York, NY}
}

@article{ouyang2025nondeterminism,
  title={An empirical study of the non-determinism of chatgpt in code generation},
  author={Ouyang, Shuyin and Zhang, Jie M and Harman, Mark and Wang, Meng},
  journal={ACM Transactions on Software Engineering and Methodology},
  volume={34},
  number={2},
  pages={1--28},
  year={2025},
  publisher={ACM New York, NY}
}

@article{li2022competition,
  title={Competition-level code generation with alphacode},
  author={Li, Yujia and Choi, David and Chung, Junyoung and Kushman, Nate and Schrittwieser, Julian and Leblond, R{\'e}mi and Eccles, Tom and Keeling, James and Gimeno, Felix and Dal Lago, Agustin and others},
  journal={Science},
  volume={378},
  number={6624},
  pages={1092--1097},
  year={2022},
  publisher={American Association for the Advancement of Science}
}

@article{li2025safegenbench,
  title={Safegenbench: A benchmark framework for security vulnerability detection in llm-generated code},
  author={Li, Xinghang and Ding, Jingzhe and Peng, Chao and Zhao, Bing and Gao, Xiang and Gao, Hongwan and Gu, Xinchen},
  journal={arXiv preprint arXiv:2506.05692},
  year={2025}
}

@inproceedings{nie2025secodeplt,
  title={{SeCodePLT}: A Unified Benchmark for Evaluating the Security Risks and Capabilities of Code {GenAI}},
  author={Nie, Yuzhou and Wang, Zhun and Yang, Yu and Jiang, Ruizhe and Tang, Yuheng and Davies, Xander and Gal, Yarin and Li, Bo and Guo, Wenbo and Song, Dawn},
  booktitle={Advances in Neural Information Processing Systems},
  volume={38},
  year={2025}
}

@misc{marjamaki2007cppcheck,
  author       = {Marjam{\"a}ki, Daniel},
  title        = {{Cppcheck}: A Tool for Static {C}/{C++} Code Analysis},
  year         = {2007},
  howpublished = {\url{http://cppcheck.sourceforge.net}},
  note         = {Accessed: December 15, 2025}
}

@misc{mitre_cwe,
  title = {Common Weakness Enumeration},
  author = {{The MITRE Corporation}},
  year = {2024},
  url = {https://cwe.mitre.org/},
  note = {Accessed: February 2026}
}

@article{llama3report2024,
  title={The {Llama} 3 Herd of Models},
  author={{Llama Team, AI@Meta}},
  journal={arXiv preprint arXiv:2407.21783},
  year={2024}
}

@article{hui2024qwen2,
      title={Qwen2. 5-Coder Technical Report},
      author={Hui, Binyuan and Yang, Jian and Cui, Zeyu and Yang, Jiaxi and Liu, Dayiheng and Zhang, Lei and Liu, Tianyu and Zhang, Jiajun and Yu, Bowen and Dang, Kai and others},
      journal={arXiv preprint arXiv:2409.12186},
      year={2024}
}

@misc{gemmateam2025gemma3technicalreport,
      title={Gemma 3 Technical Report}, 
      author={{Gemma Team, Google DeepMind}},
      year={2025},
      eprint={2503.19786},
      archivePrefix={arXiv},
      primaryClass={cs.CL},
      url={https://arxiv.org/abs/2503.19786}, 
}

@inproceedings{dai2025comprehensive,
  title={Rethinking the Evaluation of Secure Code Generation},
  author={Dai, Shih-Chieh and Xu, Jun and Tao, Guanhong},
  booktitle={Proceedings of the 48th International Conference on Software Engineering (ICSE '26)},
  year={2026},
  address={Rio de Janeiro, Brazil},
  publisher={ACM/IEEE}
}

@article{pearce2025asleep,
  title={Asleep at the keyboard? Assessing the security of {GitHub Copilot}’s code contributions},
  author={Pearce, Hammond and Ahmad, Baleegh and Tan, Benjamin and Dolan-Gavitt, Brendan and Karri, Ramesh},
  journal={Communications of the ACM},
  volume={68},
  number={2},
  pages={96--105},
  year={2025},
  publisher={ACM, New York, NY, USA}
}

@inproceedings{jain2025livecodebench,
  title={{LiveCodeBench}: Holistic and contamination free evaluation of large language models for code},
  author={Jain, Naman and Gu, Alex and Li, Wen-Ding and Yan, Fanjia and Zhang, Tianjun and Wang, Sida and Solar-Lezama, Armando and Sen, Koushik and Stoica, Ion},
  booktitle={International Conference on Learning Representations},
  year={2025}
}

@inproceedings{khoury2023secure,
  title={How Secure is Code Generated by {ChatGPT}?},
  author={Khoury, Rapha{\"e}l and Avila, Anderson R. and Brunelle, Jacob and Camara, Baba Mamadou},
  booktitle={2023 IEEE International Conference on Systems, Man, and Cybernetics (SMC)},
  pages={2445--2450},
  year={2023},
  publisher={IEEE}
}

@article{quan2025codeelo,
  title={Codeelo: Benchmarking competition-level code generation of llms with human-comparable elo ratings},
  author={Quan, Shanghaoran and Yang, Jiaxi and Yu, Bowen and Zheng, Bo and Liu, Dayiheng and Yang, An and Ren, Xuancheng and Gao, Bofei and Miao, Yibo and Feng, Yunlong and others},
  journal={arXiv preprint arXiv:2501.01257},
  year={2025}
}

@article{tony2025prompting,
  title={Prompting techniques for secure code generation: A systematic investigation},
  author={Tony, Catherine and D{\'\i}az Ferreyra, Nicol{\'a}s E and Mutas, Markus and Dhif, Salem and Scandariato, Riccardo},
  journal={ACM Transactions on Software Engineering and Methodology},
  volume={34},
  number={8},
  pages={1--53},
  year={2025},
  publisher={ACM, New York, NY}
}

@inproceedings{tony2023llmseceval,
  title={{LLMSecEval}: A Dataset of Natural Language Prompts for Security Evaluations},
  author={Tony, Catherine and Mutas, Markus and Ferrag, Mohamed Amine and Cordeiro, Lucas C.},
  booktitle={Proceedings of the 20th International Conference on Mining Software Repositories (MSR '23), Data Showcase Track},
  year={2023},
  publisher={IEEE},
  doi={10.1109/MSR59073.2023.00050}
}

@inproceedings{tihanyi2023formai,
title = {The {FormAI} Dataset: Generative {AI} in Software Security through the Lens of Formal Verification},
author = {Tihanyi, Norbert and Bisztray, Tamas and Jain, Ridhi and Ferrag, Mohamed Amine and Cordeiro, Lucas C. and Mavroeidis, Vasileios},
year = {2023},
isbn = {9798400703751},
publisher = {Association for Computing Machinery},
address = {New York, NY, USA},
url = {https://doi.org/10.1145/3617555.3617874},
doi = {10.1145/3617555.3617874},
booktitle = {Proceedings of the 19th International Conference on Predictive Models and Data Analytics in Software Engineering},
pages = {33–43},
numpages = {11},
location = {San Francisco, CA, USA},
series = {PROMISE 2023}
}

@article{tihanyi2025formaiv2,
  title={How secure is {AI}-generated code: a large-scale comparison of large language models},
  author={Tihanyi, Norbert and Bisztray, Tamas and Ferrag, Mohamed Amine and Jain, Ridhi and Cordeiro, Lucas C},
  journal={Empirical Software Engineering},
  volume={30},
  number={2},
  pages={47},
  year={2025},
  publisher={Springer}
}

@inproceedings{serebryany2012asan,
  title={{AddressSanitizer}: A Fast Address Sanity Checker},
  author={Serebryany, Konstantin and Bruening, Derek and Potapenko, Alexander and Vyukov, Dmitriy},
  booktitle={2012 USENIX Annual Technical Conference (USENIX ATC)},
  pages={309--318},
  year={2012}
}

@article{chen2021evaluating,
  title={Evaluating large language models trained on code},
  author={Chen, Mark and Tworek, Jerry and Jun, Heewoo and Yuan, Qiming and Pinto, Henrique Ponde De Oliveira and Kaplan, Jared and Edwards, Harri and Burda, Yuri and Joseph, Nicholas and Brockman, Greg and others},
  journal={arXiv preprint arXiv:2107.03374},
  year={2021}
}

@article{hendrycks2021apps,
  title={Measuring coding challenge competence with apps},
  author={Hendrycks, Dan and Basart, Steven and Kadavath, Saurav and Mazeika, Mantas and Arora, Akul and Guo, Ethan and Burns, Collin and Puranik, Samir and He, Horace and Song, Dawn and others},
  journal={arXiv preprint arXiv:2105.09938},
  year={2021}
}

@inproceedings{siddiq2022securityeval,
  title={SecurityEval dataset: mining vulnerability examples to evaluate machine learning-based code generation techniques},
  author={Siddiq, Mohammed Latif and Santos, Joanna CS},
  booktitle={Proceedings of the 1st international workshop on mining software repositories applications for privacy and security},
  pages={29--33},
  year={2022}
}

@inproceedings{siddiq2022codesmells,
  title={An Empirical Study of Code Smells in Transformer-based Code Generation Techniques},
  author={Siddiq, Mohammed Latif and Santos, Joanna C. S.},
  booktitle={Proceedings of the 22nd {IEEE} International Working Conference on Source Code Analysis and Manipulation (SCAM '22)},
  pages={71--82},
  year={2022},
  publisher={IEEE},
  doi={10.1109/SCAM55253.2022.00015}
}

@inproceedings{fang2024lmcbench,
  title={Large language models for code analysis: Do $\{$LLMs$\}$ really do their job?},
  author={Fang, Chongzhou and Miao, Ning and Srivastav, Shaurya and Liu, Jialin and Zhang, Ruoyu and Fang, Ruijie and Tsang, Ryan and Nazari, Najmeh and Wang, Han and Homayoun, Houman and others},
  booktitle={33rd USENIX Security Symposium (USENIX Security 24)},
  pages={829--846},
  year={2024}
}

@article{jiang2026survey,
  title={A survey on large language models for code generation},
  author={Jiang, Juyong and Wang, Fan and Shen, Jiasi and Kim, Sungju and Kim, Sunghun},
  journal={ACM Transactions on Software Engineering and Methodology},
  volume={35},
  number={2},
  pages={1--72},
  year={2026},
  publisher={ACM New York, NY}
}

@inproceedings{lattner2004llvm,
  author    = {Lattner, Chris and Adve, Vikram},
  title     = {{LLVM}: A Compilation Framework for Lifelong Program Analysis \& Transformation},
  booktitle = {Proceedings of the International Symposium on Code Generation and Optimization (CGO)},
  pages     = {75--86},
  year      = {2004},
  publisher = {IEEE Computer Society}
}

@inproceedings{gadelha2018esbmc,
  title={{ESBMC} 5.0: An Industrial-Strength {C} Model Checker},
  author={Gadelha, Mikhail R. and Monteiro, Felipe R. and Morse, Jeremy and Cordeiro, Lucas C. and Fischer, Bernd and Nicole, Denis A.},
  booktitle={Proceedings of the 33rd ACM/IEEE International Conference on Automated Software Engineering (ASE)},
  pages={888--891},
  year={2018},
  publisher={ACM},
  doi={10.1145/3238147.3240481}
}

@inproceedings{gadelha2019esbmc_industrial,
  title={ESBMC v6. 0: Verifying C Programs Using k-Induction and Invariant Inference: (Competition Contribution)},
  author={Gadelha, Mikhail R and Monteiro, Felipe and Cordeiro, Lucas and Nicole, Denis},
  booktitle={International Conference on Tools and Algorithms for the Construction and Analysis of Systems},
  pages={209--213},
  year={2019},
  organization={Springer}
}

@inproceedings{kroening2014cbmc,
  title={{CBMC}--{C} Bounded Model Checker: (Competition Contribution)},
  author={Kroening, Daniel and Tautschnig, Michael},
  booktitle={International Conference on Tools and Algorithms for the Construction and Analysis of Systems},
  pages={389--391},
  year={2014},
  organization={Springer}
}

@inproceedings{biere1999bounded,
  title={Symbolic model checking without {BDDs}},
  author={Biere, Armin and Cimatti, Alessandro and Clarke, Edmund and Zhu, Yunshan},
  booktitle={International conference on tools and algorithms for the construction and analysis of systems},
  pages={193--207},
  year={1999},
  organization={Springer}
}

@inproceedings{szekeres2013sok,
  title={{SoK}: Eternal War in Memory},
  author={Szekeres, L{\'a}szl{\'o} and Payer, Mathias and Wei, Tao and Song, Dawn},
  booktitle={2013 IEEE Symposium on Security and Privacy (S\&P)},
  pages={48--62},
  year={2013},
  publisher={IEEE},
  doi={10.1109/SP.2013.13}
}

@inproceedings{perry2023users,
  title={Do Users Write More Insecure Code with {AI} Assistants?},
  author={Perry, Neil and Srivastava, Megha and Kumar, Deepak and Boneh, Dan},
  booktitle={Proceedings of the 2023 ACM SIGSAC Conference on Computer and
             Communications Security (CCS '23)},
  pages={2785--2799},
  year={2023},
  publisher={ACM},
  doi={10.1145/3576915.3623157}
}

@inproceedings{sandoval2023lost,
  title={Lost at {C}: A User Study on the Security Implications of Large Language
         Model Code Assistants},
  author={Sandoval, Gustavo and Pearce, Hammond and Nys, Teo and Karri, Ramesh
          and Garg, Siddharth and Dolan-Gavitt, Brendan},
  booktitle={32nd USENIX Security Symposium (USENIX Security '23)},
  pages={2205--2222},
  year={2023},
  publisher={USENIX Association}
}

@article{asare2023github,
  title={Is github’s copilot as bad as humans at introducing vulnerabilities in code?},
  author={Asare, Owura and Nagappan, Meiyappan and Asokan, Nirmal},
  journal={Empirical Software Engineering},
  volume={28},
  number={6},
  pages={129},
  year={2023},
  publisher={Springer}
}

@inproceedings{moura2008z3,
  title={{Z3}: An efficient {SMT} solver},
  author={De Moura, Leonardo and Bj{\o}rner, Nikolaj},
  booktitle={International conference on Tools and Algorithms for the Construction and Analysis of Systems},
  pages={337--340},
  year={2008},
  organization={Springer}
}

@article{openai2023gpt4,
  title={{GPT-4} Technical Report},
  author={{OpenAI}},
  journal={arXiv preprint arXiv:2303.08774},
  year={2023}
}

@article{roziere2023code,
  title={Code llama: Open foundation models for code},
  author={Roziere, Baptiste and Gehring, Jonas and Gloeckle, Fabian and Sootla, Sten and Gat, Itai and Tan, Xiaoqing Ellen and Adi, Yossi and Liu, Jingyu and Sauvestre, Romain and Remez, Tal and others},
  journal={arXiv preprint arXiv:2308.12950},
  year={2023}
}

@article{austin2021program,
  title={Program synthesis with large language models},
  author={Austin, Jacob and Odena, Augustus and Nye, Maxwell and Bosma, Maarten and Michalewski, Henryk and Dohan, David and Jiang, Ellen and Cai, Carrie and Terry, Michael and Le, Quoc and others},
  journal={arXiv preprint arXiv:2108.07732},
  year={2021}
}

@article{bessey2010fewbillion,
  title={A Few Billion Lines of Code Later: Using Static Analysis to Find Bugs
         in the Real World},
  author={Bessey, Al and Block, Ken and Chelf, Ben and Chou, Andy and Fulton, Bryan
          and Hallem, Seth and Henri-Gros, Charles and Kamsky, Asya and McPeak, Scott
          and Engler, Dawson},
  journal={Communications of the {ACM}},
  volume={53},
  number={2},
  pages={66--75},
  year={2010},
  publisher={ACM},
  doi={10.1145/1646353.1646374}
}

@inproceedings{beyer2024svcomp,
  title={State of the art in software verification and witness validation: {SV-COMP} 2024},
  author={Beyer, Dirk},
  booktitle={International Conference on Tools and Algorithms for the Construction and Analysis of Systems},
  pages={299--329},
  year={2024},
  organization={Springer}
}

@article{li2015techdebt_mapping,
  title={A systematic mapping study on technical debt and its management},
  author={Li, Zengyang and Avgeriou, Paris and Liang, Peng},
  journal={Journal of systems and software},
  volume={101},
  pages={193--220},
  year={2015},
  publisher={Elsevier}
}

@misc{openrouter,
  title = {{OpenRouter}: A Unified Interface for {LLMs}},
  author = {{OpenRouter}},
  year = {2024},
  url = {https://openrouter.ai/}
}

@inproceedings{fan2020ac,
  title={{AC/C++} code vulnerability dataset with code changes and {CVE} summaries},
  author={Fan, Jiahao and Li, Yi and Wang, Shaohua and Nguyen, Tien N},
  booktitle={Proceedings of the 17th international conference on mining software repositories},
  pages={508--512},
  year={2020}
}

@article{bhatt2023purple,
  title={Purple llama cyberseceval: A secure coding benchmark for language models},
  author={Bhatt, Manish and Chennabasappa, Sahana and Nikolaidis, Cyrus and Wan, Shengye and Evtimov, Ivan and Gabi, Dominik and Song, Daniel and Ahmad, Faizan and Aschermann, Cornelius and Fontana, Lorenzo and others},
  journal={arXiv preprint arXiv:2312.04724},
  year={2023}
}

@inproceedings{peng2025cweval,
  title={Cweval: Outcome-driven evaluation on functionality and security of llm code generation},
  author={Peng, Jinjun and Cui, Leyi and Huang, Kele and Yang, Junfeng and Ray, Baishakhi},
  booktitle={2025 IEEE/ACM International Workshop on Large Language Models for Code (LLM4Code)},
  pages={33--40},
  year={2025},
  organization={IEEE}
}

@article{vero2025baxbench,
  title={{BaxBench}: Can {LLMs} generate correct and secure backends?},
  author={Vero, Mark and M{\"u}ndler, Niels and Chibotaru, Victor and Raychev, Veselin and Baader, Maximilian and Jovanovi{\'c}, Nikola and He, Jingxuan and Vechev, Martin},
  journal={arXiv preprint arXiv:2502.11844},
  year={2025}
}

@inproceedings{zhang2024seccoder,
  title={Seccoder: Towards generalizable and robust secure code generation},
  author={Zhang, Boyu and Du, Tianyu and Tong, Junkai and Zhang, Xuhong and Chow, Kingsum and Cheng, Sheng and Wang, Xun and Yin, Jianwei},
  booktitle={Proceedings of the 2024 Conference on Empirical Methods in Natural Language Processing},
  pages={14557--14571},
  year={2024}
}

@misc{codeforces2024airule,
  author       = {{Codeforces}},
  title        = {Rule Restricting the Use of {AI}},
  year         = {2024},
  howpublished = {Codeforces Blog, Entry 133941},
  url          = {https://codeforces.com/blog/entry/133941},
  note         = {Accessed: 2026-11-16}
}

@misc{clang2026ubsan,
  author       = {{LLVM Project}},
  title        = {{UndefinedBehaviorSanitizer}},
  howpublished = {Clang documentation},
  url          = {https://clang.llvm.org/docs/UndefinedBehaviorSanitizer.html},
  note         = {Accessed: 2026-11-18}
}

\end{document}